\documentclass[aoas]{imsart}

\RequirePackage{amsthm,amsmath,amsfonts,amssymb}
\RequirePackage[authoryear]{natbib}
\RequirePackage[colorlinks,citecolor=blue,urlcolor=blue]{hyperref}
\RequirePackage{graphicx}

\RequirePackage[nolist]{acronym}
\usepackage{caption}
\usepackage{subcaption}
\usepackage[boxed]{algorithm2e}
\usepackage{tikz}
\definecolor{transparent}{RGB}{255,255,255}
\usetikzlibrary{intersections,shapes,arrows, patterns}
\tikzstyle{state} = [circle, minimum width=.9cm, minimum height=.9cm, draw=blue, text centered]
\tikzstyle{obs} = [circle, minimum width=.9cm, minimum height=.9cm, draw=blue, text centered, fill=lightgray]
\tikzstyle{plate} = [rectangle, minimum width=5cm, minimum height=5.5cm, text centered, draw=black, fill=white]
\tikzstyle{namecompartment} = [rectangle, minimum width=0.5cm, minimum height=0.5cm, text centered ]
\tikzstyle{compartment} = [rectangle, minimum width=1cm, minimum height=1cm, text centered, draw=blue, fill=white]
\tikzstyle{comp} = [rectangle, minimum width=.75cm, minimum height=.75cm, text centered, draw=blue, fill=white]

\newcommand{\rev}[1]{{\color{black}#1}}
\newcommand{\vtheta}{\boldsymbol{\theta}}

\startlocaldefs

\theoremstyle{remark}

\endlocaldefs

\begin{document}

\begin{frontmatter}
\title{Inferring epidemics from multiple dependent data via pseudo-marginal methods}
\runtitle{Epidemic inference with dependent data}

\begin{aug}
\author[A,B]{\fnms{Alice} \snm{Corbella}\ead[label=e1, mark]{alice.corbella@warwick.ac.uk}},
\author[B]{\fnms{Anne M} \snm{Presanis}\ead[label=e2,mark]{anne.presanis@mrc-bsu.cam.ac.uk}}
\author[C,B]{\fnms{Paul J} \snm{Birrell}\ead[label=e3,mark]{paul.birrell@ukhsa.gov.uk}}
\and
\author[B,C]{\fnms{Daniela} \snm{De Angelis}\ead[label=e4,mark]{daniela.deangelis@mrc-bsu.cam.ac.uk}}
\address[A]{Department of Statistics,
University of Warwick,
\printead{e1}}

\address[B]{MRC Biostatistics Unit,
University of Cambridge,
\printead{e2,e4}}
\address[C]{UK Health Security Agency
\printead{e3}}
\end{aug}

\begin{abstract}
Health-policy planning requires evidence on the burden that epidemics place on healthcare systems. Multiple, often dependent, datasets provide a noisy and fragmented signal from the unobserved epidemic process including transmission and severity dynamics. 

This paper explores important challenges to the use of state-space models for epidemic inference when multiple dependent datasets are analysed. We propose a new semi-stochastic model that exploits deterministic approximations for large-scale transmission dynamics while retaining stochasticity in the occurrence and reporting of relatively rare severe events. This model is suitable for many real-time situations including large seasonal epidemics and pandemics. 
Within this context, we develop algorithms to provide exact parameter inference and test them via simulation. 

Finally, we apply our joint model and the proposed algorithm to several surveillance data on the 2017-18 influenza epidemic in England to reconstruct transmission dynamics and estimate  the daily new influenza infections as well as severity indicators such as the case-hospitalisation risk and the hospital-intensive care risk. 

\end{abstract}

\end{frontmatter}

\section{Introduction}\rev{The outbreak of an epidemic and the generation of cases at different levels of severity is a complex dynamical system involving many layers of randomness. Multiple datasets on the related processes might be available at various stages of the epidemic, mainly in the form of time series, but they tend to be affected by noise, delays and interdependencies. One example of an epidemic with multiple data sources is influenza. In the UK, some mild cases of influenza, not confirmed by laboratory test, are registered upon their \ac{GP} visit, while confirmed severe cases are recorded when admitted to hospital or \ac{IC} unit. Estimating transmission and severity dynamics from these data is key for health policy planning but it can prove challenging from an inferential perspective: it is not straightforward to use information from different data sources to inform a single model capturing disease spread, development of symptoms, access to health care and case ascertainment. A unifying modelling framework combined with a set of robust inferential tools could help to provide a comprehensive picture of several aspects of an outbreak.} 

\Acp{SSM} have often been used to model the spread of a viral infection in a population, since they provide a practical and simple framework to describe the generation of epidemic time series \citep[e.g.][]{dureau2013ssm}. 
\acp{SSM} are latent variable representations of dynamical systems evolving over specific domains (e.g. time, space, generations) \citep{schon2018probabilistic}. The definition of an \ac{SSM} involves the specification of a system of stochastic or deterministic equations that define the behaviour of the latent process, $ X_t $, where $ t $ indexes the domain (e.g. \textit{time}), as well as the emission of observable quantities throughout the domain, denoted by $ Y_t $ \citep{birrell2018evidence}. Thanks to their flexibility, \ac{SSM} representations of observable phenomena have been widely used in many fields, from indoor positioning problems in engineering \citep{solin2018modeling}, to environmental studies \citep{anderson1996method}, to epidemic models \citep{magpantay2016pertussis}. 

In modelling epidemics as \acp{SSM}, the latent process generally accounts for: (i) a transmission process, describing the spread of infection through a population, (ii) a severity process, modelling the likelihood of infected individuals to become severe cases (e.g. symptomatic, hospitalised, etc.), and (iii) a detection process that links processes (i) and (ii) to available observations. In addition,  other sources of randomness might affect the generation of epidemic data, including seasonal fluctuations in parameters and misclassification of cases unrelated to the pathogen of interest. 

\rev{The formulation of such a \ac{SSM} to describe epidemic data, requires several model choices to make. A key choice is the level of stochasticity assumed in the system. Wholly deterministic systems have been shown to be a reasonable approximation to epidemic dynamics under specific conditions (e.g. when the number of new infections is sufficiently large). 

Such models can include complex dynamical components: }\cite{birrell2011bayesian}, for example, assumes that an age-structured deterministic discrete-time system could approximate the spread of the 2009 pandemic influenza; it accounts for delays between infection and detection of symptomatic cases and places most of the randomness in the observation model. A similar approach, with more complex layers, is considered in other works including \cite{weidemann2014bayesian}, analysing Rotavirus data, and in recent works analysing COVID-19 data, e.g. \cite{keeling2020fitting} and \cite{birrell2021real}. 
\rev{However, models adopting higher levels of stochasticity can more realistically account for the noise and variability in the data, but they often rely on simplifying assumptions on the dynamics.} An example is the work of \cite{lekone2006statistical}: here a discrete-time stochastic model is fitted to Ebola data and substantial, sometimes unrealistic, assumptions are made, including exponentially distributed time to events and full ascertainment of the cases. These kinds of models have been particularly utilised for small-size epidemics where the large-numbers deterministic approximation does not hold (e.g. \cite{nishiura2011real}, \cite{funk2018real}). 

Hybrid models that exploit the strength from both approaches could potentially prove useful in many contexts. The transmission dynamics of large seasonal epidemics and of pandemics can be reliably approximated by deterministic systems, which, compared to their correspondent stochastic versions, allow greater model complexity (mixing heterogeneity, many compartments, etc.). In many contexts, however, the number of cases that become severe and detected might be smaller and subject to significant stochastic variation. Stochastic systems would be required to describe the severity process, marking a difference from many models listed above. 

While a model with wholly stochastic relationships except for the deterministic transmission dynamics has greater realism, it might be more challenging from an inferential perspective, especially when multiple data sources are considered. If individuals are monitored at different levels of severity in multiple datasets, a Bayesian evidence synthesis framework \citep{presanis2013conflict, presanis2014synthesising} should be employed, where correlations and delays among these layers of severity (and their related data sources) are accounted for.

This paper addresses the problem of inference of epidemics from multiple, possibly dependent, data by presenting a general framework for the inferential problem; it identifies and tackles the challenges that arise in this context, considering the specific case of a deterministic transmission model with multiple layers of stochasticity in the severity and detection stage. 
Section \ref{sec2} provides a background to the methods used for modelling and inference, specifically \acp{SSM} and pseudo-marginal methods. Section \ref{sec3} considers their application to the epidemic setting, specifying a model that mirrors realistic situations, presents challenges of such models and proposes an inferential routine. The aim of Section \ref{sec4} is to compare the model presented in Section \ref{sec3} to a similar model that, through a simplifying assumption, does not account for the intrinsic dependence between datasets, inducing under-estimation of parameter uncertainty. Lastly Section \ref{sec5} presents the analysis of real data on cases of seasonal influenza at different levels of severity, appropriately attributing stochasticity to each process.  


\section{State-space models and their inference}\label{sec2}


\subsection{State-space models}
\rev{A \ac{SSM} has two components: a latent process, $\left\lbrace {X_t} \right\rbrace _{t\geq0} $ representing the underlying states
; and an observed process $\left\lbrace {Y_t} \right\rbrace _{t\geq1} $. Here, the domain of the process is assumed to be discrete time, with intervals indexed by $ t $. 
A 
\ac{SSM} is defined through the state
and the observation equations as follows: 
\begin{eqnarray}
\label{eq1_1}X_0|\vtheta &\sim&  p(x_0|\vtheta)\\
\label{eq1_2}{X_t}|({X_{t-1}}, \vtheta) &\sim& p({x_t}|{x_{t-1}},\vtheta)\\
\label{eq1_3}Y_t|({X_{t}}, \vtheta) &\sim & p(y_t|{x_{t}},\vtheta)
\end{eqnarray}
for  $ t=1, 2, \dots, T $. These equations express the joint distribution of the initial state of the system, $ X_0 $, of the latent states, $ X_t$, and of the observations, $ Y_t $,  parameterised by a vector $ \vtheta $ \citep{brockwell2016introduction}. Equations \eqref{eq1_1} to \eqref{eq1_3} define a Markovian 
\ac{SSM} 
\citep{king2015statistical} 
 with} the following full probability model:
\begin{equation}
p({x_{0:T}}, y_{1:T}|{\vtheta})= \prod_{t=1}^{T}p(y_t|{x_t}, {\vtheta})\prod_{t=1}^{T}p({x_t}|{x_{t-1}}, {\vtheta}) p({x_0}| {\vtheta}) 
\label{eq2}
\end{equation}
where, through Markovianity and conditional independence, the joint density can be factorised into the state and observation densities.

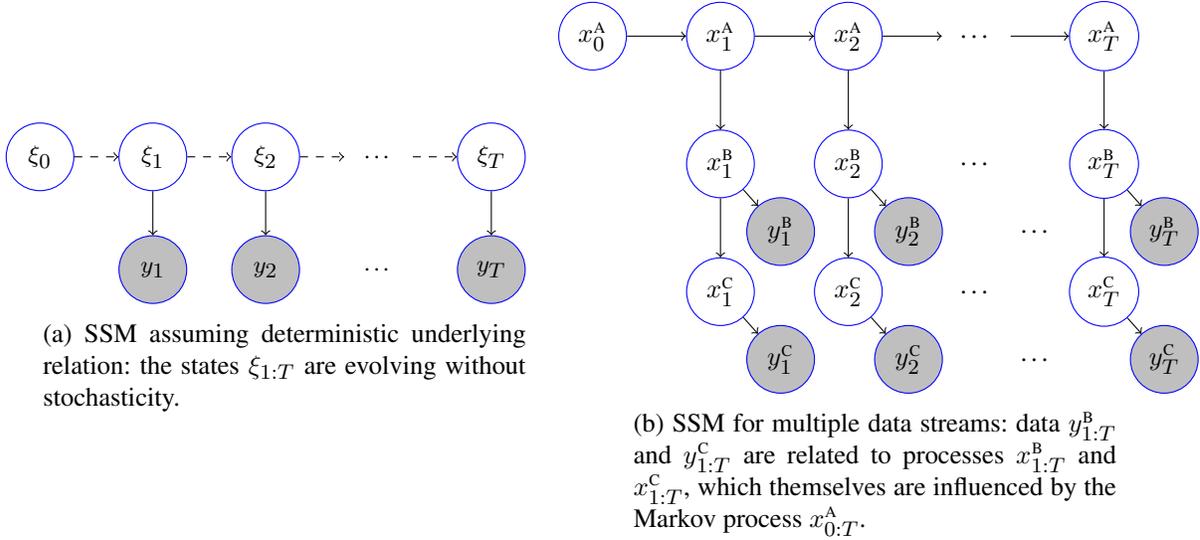
\begin{figure}[h]
\begin{subfigure}[c]{6.42cm}
\centering
\hspace{-0.5cm}\begin{tikzpicture} [node distance=1.5cm] 
\node(x0)[state]{$ x_0 $};
\node(x1)[state, right of=x0]{$ x_1 $};
\node(x2)[state, right of=x1]{$ x_2 $};
\node(xn)[state, right of=x2, draw=transparent]{\dots};
\node(xT)[state, right of=xn]{$ x_T $};
\draw [->] (x0) -- (x1);
\draw [->] (x1) -- (x2);
\draw [->] (x2) -- (xn);
\draw [->] (xn) -- (xT);
\node(y0)[obs, below of=x0, fill=transparent, draw=transparent]{ };
\node(y1)[obs, right of=y0]{$ y_1 $};
\node(y2)[obs, right of=y1]{$ y_2 $};
\node(yn)[obs, right of=y2, fill=transparent, draw=transparent]{\dots};
\node(yT)[obs, right of=yn]{$ y_T $};
\draw [->] (x1) -- (y1);
\draw [->] (x2) -- (y2);
\draw [->] (xT) -- (yT);
\end{tikzpicture}
\caption{Simple SSM evolving over time: the state $ x_{1:T} $ is unknown and only signal $y_{1:T}$ is observed. }
\label{fig2b}
\end{subfigure}\hfill
\begin{subfigure}[c]{6.42cm}
\centering
\hspace*{-1cm}\begin{tikzpicture} [node distance=1.7cm] \small
\node(x0)[state]{$ x^\textsc{a}_0 $};
\node(x1)[state, right of=x0]{$x^\textsc{a}_1 $};
\node(x2)[state, right of=x1]{$ x^\textsc{a}_2 $};
\node(xn)[state, right of=x2, draw=transparent]{\dots};
\node(xT)[state, right of=xn]{$ x^\textsc{a}_T $};
\draw [->, dashed] (x0) -- (x1);
\draw [->, dashed] (x1) -- (x2);
\draw [->, dashed] (x2) -- (xn);
\draw [->, dashed] (xn) -- (xT);
\node(xb0)[state, below of=x0, fill=transparent, draw=transparent]{ };
\node(xb1)[state, right of=xb0]{$ x^\textsc{b}_1 $};
\node(xb2)[state, right of=xb1]{$ x^\textsc{b}_2 $};
\node(xbn)[state, right of=xb2, fill=transparent, draw=transparent]{\dots};
\node(xbT)[state, right of=xbn]{$ x^\textsc{b}_T $};
\draw [->] (x1) -- (xb1);
\draw [->] (x2) -- (xb2);
\draw [->] (xT) -- (xbT);
\node(yb1)[obs, below of=xb1, xshift=0.8cm, yshift=0.8cm]{$ y^\textsc{b}_1 $};
\node(yb2)[obs, right of=yb1]{$ y^\textsc{b}_2 $};
\node(ybn)[obs, right of=yb2, fill=transparent, draw=transparent]{\dots};
\node(ybT)[obs, right of=ybn]{$ y^\textsc{b}_T $};
\draw [->] (xb1) -- (yb1);
\draw [->] (xb2) -- (yb2);
\draw [->] (xbT) -- (ybT);
\node(xc0)[state, below of=xb0, fill=transparent, draw=transparent]{ };
\node(xc1)[state, right of=xc0]{$ x^\textsc{c}_1 $};
\node(xc2)[state, right of=xc1]{$ x^\textsc{c}_2 $};
\node(xcn)[state, right of=xc2, fill=transparent, draw=transparent]{\dots};
\node(xcT)[state, right of=xcn]{$ x^\textsc{c}_T $};
\draw [->] (xb1) -- (xc1);
\draw [->] (xb2) -- (xc2);
\draw [->] (xbT) -- (xcT);
\node(yc1)[obs, below of=xc1, xshift=0.8cm, yshift=0.8cm]{$ y^\textsc{c}_1 $};
\node(yc2)[obs, right of=yc1]{$ y^\textsc{c}_2 $};
\node(ycn)[obs, right of=yc2, fill=transparent, draw=transparent]{\dots};
\node(ycT)[obs, right of=ycn]{$ y^\textsc{c}_T $};
\draw [->] (xc1) -- (yc1);
\draw [->] (xc2) -- (yc2);
\draw [->] (xcT) -- (ycT);
\end{tikzpicture}
\caption{SSM for multiple data streams: data $ y^\textsc{b}_{1:T} $ and $ y^\textsc{c}_{1:T} $ are related to processes $ x^\textsc{b}_{1:T} $ and $ x^\textsc{c}_{1:T} $, which themselves are related to the deterministic process $ x^\textsc{a}_{0:T} $.}
\label{fig2a}
\end{subfigure}
\caption{Some examples of graphical models for \acp{SSM}. Grey nodes correspond to observed variables and white nodes are latent variables. Solid arrows express stochastic dependence among \acp{r.v.}, while dashed arrows express deterministic links.}
\label{f2}
\end{figure}

A \ac{SSM} can be represented as a graphical model where a graph $ \mathcal{G}=(\mathcal{V}, \mathcal{E}) $ encodes the conditional independence structure (edges $  \mathcal{E} $) between \acp{r.v.} (nodes $ \mathcal{V} $). 
See Figure \ref{f2} for an illustration of the state and observational processes of an \ac{SSM} where the dependence on the parameter $ \vtheta $ is omitted for simplicity.

This framework could encompass many different modelling assumptions, \rev{ranging from the basic \ac{SSM} of figure \ref{fig2a} to multidimensional state processes with deterministic and stochatsic  relations, as exemplified in Figure \ref{fig2b}}.

\subsection{Inference in state-space models}\label{sec2.3}
\rev{Observations $\left\lbrace {y_t} \right\rbrace _{t\geq1} $ can be used to infer the latent process $\left\lbrace {X_t} \right\rbrace _{t\geq0} $ (state inference), typically conditional on specific values of the static parameter $ \vtheta $, or to estimate the static parameter $\vtheta  $ (parameter inference), usually through the marginal distribution from the joint posterior density for $\left(\vtheta, x_{0:t}\right)$ \citep{lindsten2013backward}}. 

\rev{State inference relies on the estimation of the filtering distribution,  $ p(x_{t}|y_{1:t}, \vtheta ) $, for $ t=1, \dots, T $, which is the state distribution at time $ t $, conditional on data and parameters.

The filtering distribution 
can often be approximated recursively by a two-step procedure that alternates approximation of the state distribution given the observations (measurement update) and given the previous states (prediction update).
This allows the approximation of the likelihood of the data $ y_{1:T} $, given a parameter value $ \vtheta $ and marginally w.r.t. the state distribution $ X_{1:T} $ in Equation \eqref{eq5}: }
\begin{equation}\label{eq5}
p(y_{1:T}|\vtheta) = \prod_{t=1}^{T} p(y_t|y_{1:t-1}, \vtheta) = \prod_{t=1}^{T} \int_{X_{0:t}}  p(y_t, x_{0:t}|y_{1:t-1}, \vtheta) \text{ d}x_{0:t}
\end{equation}
\rev{The most common among these simulation-based procedures is the \ac{BPF} \citep{arulampalam2002tutorial}, which 
provides an unbiased estimator $ \widehat{p}(y_{1:T}|\vtheta)  $ of the likelihood of the data given a parameter value $ \vtheta $, although with high computational cost for large $T$. 

}

\rev{This approximated likelihood can be used to drive inference of the parameter $\vtheta$ \citep{andrieu2010particle}. }
\rev{Within the Bayesian framework, assumed here to enable a complex synthesis of available evidence, 
{pseudo-marginal methods} (\citet{andrieu2009pseudo}), provide a simple way to integrate simulation-based approximation of the likelihood (such as the \ac{BPF}) into \ac{MCMC} algorithms for parameter inference by replacing the likelihood in the acceptance rate of the \ac{MH} algorithm with its approximation.}
\rev{
Two such pseudo-marginal algorithms are employed throughout this paper: \ac{GIMH} \citep{beaumont2003estimation} and \ac{MCWM} \citep{andrieu2009pseudo}\rev{, see Supplementary Information for further details.
	.}}

\subsection{State-space models for epidemics}
The \ac{SSM} \rev{framework} marries well with epidemic models: available data are only a partially-observed signal of latent variables which encapsulate all the data generative process (e.g. transmission, severity, background noise). 
More specifically, the transmission model component usually consists of stochastic or deterministic system of equations that describe the flow of individuals between disjoint compartments according to their infection status, { e.g.} susceptible, infectious, recovered and immune.

\rev{Deterministic or stochastic transmission models have been used from the earliest works on disease dynamics \citep{kermack1927contribution}: the former allowing for more-complex systems but approximating layers of stochasticity \citep{keeling2011modeling}, and the latter heavily relying on simplifying model assumptions to derive closed forms of the likelihood \citep{britton2010stochastic, andersson2012stochastic}. }


In recent decades \rev{the \ac{SSM} framework} has been used for epidemic dynamics more explicitly \rev{\citep{breto2009time, dukic2012tracking, dureau2013ssm, mckinley2014simulation, shubin2016revealing} and has become increasingly popular during the recent COVID-19 pandemic \citep{read2021novel,knock2021key,  pooley2022estimation}. } \rev{Although} \ac{SSM} inference is possible, many simplifying assumptions are needed to allow computation, particularly under time constraints. These simplifications often include assuming independence between data stream\rev{s}, dropping transmission compartments, assuming full ascertainment of the cases, and discretising time in large intervals. Importantly, some but not all parts of the state process of these models involve large numbers, and could be be reliably approximated by its deterministic counterpart.

\rev{
\subsection{A semi-stochastic model and its inference}
Within the application of \acp{SSM} to epidemics, there seem to be a lack of hybrid models that exploit the deterministic approximations of transmission dynamics appropriate for large epidemics and pandemics while properly accounting for stochasticity in the latent states whose dynamics cannot be deterministically approximated, e.g. the states that involve severe cases. In what follows we propose and discuss such a semi-stochastic model
. More specifically, we propose for the first time a deterministic transmission model coupled with a stochastic severity model, accounting for stochasticity only in those states where the counts are very small (e.g. hospitalised patients). 
From an inferential point of view, the fitting of such an hybrid model could be accommodated within standard \ac{SSM} settings with time domain. Nevertheless, differently from the typical Markovian \acp{SSM}, the introduction of the deterministic dynamics eliminates the dependence of the hidden states across time, which is the usual bottleneck of sequential methods, allowing a convenient trade-off between model realism and computational burden.}

\section{Multiple data on epidemics}\label{sec3}
\rev {When an infectious disease spreads in a population, multiple signals of its presence over time might be available in the form of time-series counts. Each of these time series  is related to a specific level of severity (e.g. mild symptomatic cases, patients that require hospitalisation) and could be affected by specific sources of noise and, possibly, bias. However, all data are linked to the underlying transmission process. Therefore, the joint analysis of multiple data is likely to provide a better understanding of the infection and severity process than separate analyses of single datasets. This joint analysis, however, introduces several challenges.}

\rev{In what follows
the unknown model states, denoted by the letter $ X $, represent counts of individuals experiencing specific events at specific time points. The upper index of $ X $ identifies the \textit{type of event} and the lower index identifies the \textit{time} at which the event takes place (e.g. $ X^\textsc{h}_t $ is the number of people \underline{$ \textsc{h} $}ospitalised, at time $ t $). Conditional events  e.g. the number of people \underline{$ \textsc{h} $}ospitalised, at $ t $ that had \underline{$ \textsc{v} $}iral infection at time $ s $, are denoted as $X_{t}^{\textsc{h}|\textsc{v}_{s}}$. When states are deterministic functions of static parameters, the $ X $ is replaced by $ \xi $ (e.g. $ \xi^\textsc{v}_t $ is the number of people that had a \underline{$ \textsc{v} $}iral infection, at $ t $), see Figure \ref{fig2a}.  

A similar indexing system is used for conditional probabilities of severe events  $ \rho $ (e.g. $  \rho^{\textsc{h}|\textsc{v}} $ is the probability of \underline{$ \textsc{h} $}ospitalisation, given \underline{$ \textsc{v} $}iral infection) and for the observations  $ y $ (e.g. $ y^\textsc{h}_t $ is the number of observed \underline{$ \textsc{h} $}ospitalisations, at $ t $)}.

\subsection{Dependency between data streams at time $t$}
Let $ X^\textsc{v}_t $ be the number of new viral infections at time $ t $ of a specific epidemic in a given population. Infected individuals could experience a series of increasingly severe events  \citep{presanis2009severity}, e.g. they might be hospitalised \rev{($ \textsc{h} $)}, require intensive care \rev{($ \textsc{ic} $)}, die \rev{($ \textsc{d} $)}; \rev{with more severe cases being a subset of cases with milder disease}, as represented in Figure \ref{fig3a}. 
\rev{The severity process is here assumed to be linked directly to the number of new infections (i.e. those moving to the $E$xposed state). These newly infected individuals enter the severity-process pathway (Figure \ref{fig3a}), in addition to the transmission pathway (Figure \ref{fig3c}) which they had started when infected. }

\begin{figure}[h]
\begin{subfigure}[c]{6.42cm}
\centering
\begin{tikzpicture}
\coordinate (A) at (-2.5,0) {};
\coordinate (B) at ( 2.5,0) {};
\coordinate (C) at (0,4) {};
\draw[name path=AC, blue] (A) -- (C);
\draw[name path=BC, blue] (B) -- (C);
\foreach \y/\A in {
0/$ \textsc{v} $,
1/ $ \textsc{h} $,
2/$ \textsc{ic} $, 
3/$ \textsc{d} $      } {
\path[name path=horiz, blue] (A|-0,\y) -- (B|-0,\y);
\draw[name intersections={of=AC and horiz,by=P},
name intersections={of=BC and horiz,by=Q}, blue] (P) -- (Q)
node[midway,above,align=center,text width=
\dimexpr(3.5em-\y em)*5\relax, black] {\A};
}
\end{tikzpicture}
\caption{Pyramid representation of the severity pathway.}
\label{fig3a}
\end{subfigure}
\hspace*{1cm}	\begin{subfigure}[c]{6.42cm}
\centering
\begin{tikzpicture}[node distance=1.3cm]
\node (PL)[plate, xshift=-0.1cm , yshift=2.9cm]{};
\node (PL)[plate, xshift=-0.05cm , yshift=2.85cm]{};
\node (PL)[plate, yshift=2.8cm]{ };
\node (mu)[state, yshift=5cm, xshift=0cm] {$ x^\textsc{v}_{t}$};
\node (xh)[state, below of=mu, xshift=-1cm] {$ x^\textsc{h}_{t}$};
\node (xic)[state, below of=xh] {$ x^\textsc{ic}_{t}$};
\node (xl)[state, below of=xic] {$ x^\textsc{d}_{t}$};
\node (yic)[obs, below of=xic, xshift=2cm, yshift=1cm] {$ y^\textsc{ic}_{t}$};
\node (yh)[obs, below of=xh, xshift=2cm, yshift=1cm] {$ y^\textsc{h}_{t}$};
\node (yl)[obs, below of=xl, xshift=2cm, yshift=1cm] {$ y^\textsc{d}_{t}$};
\node (phiH)[state, left of=xh, xshift=-1cm, yshift=1cm] {$ \rho^{\textsc{h}|\textsc{v}}$};
\node (phiIC)[state, left of=xic, xshift=-1cm, yshift=1cm] {$ \rho^{\textsc{ic}|\textsc{h}}$};
\node (phil)[state, left of=xl, xshift=-1cm, yshift=1cm] {$ \rho^{\textsc{d}|\textsc{ic}}$};
\node (xiH)[state, right of=yh,  xshift=1cm, yshift=1cm] {$ \zeta^\textsc{h}$};
\node (xiIC)[state, right of=yic,  xshift=1cm, yshift=1cm] {$ \zeta^\textsc{ic}$};
\node (xil)[state, right of=yl,  xshift=1cm, yshift=1cm] {$ \zeta^\textsc{d}$};
\draw [->, gray] (phiIC) -- node[anchor=south] { } (xic);
\draw [->, gray] (phiH) -- node[anchor=south] { } (xh);
\draw [->, gray] (phil) -- node[anchor=south] { } (xl);
\draw [->, gray] (xiH) -- node[anchor=south] { } (yh);
\draw [->, gray] (xiIC) -- node[anchor=south] { } (yic);
\draw [->, gray] (xil) -- node[anchor=south] { } (yl);
\draw [->, gray] (xh) -- node[anchor=south] { } (yh);
\draw [->, gray] (xic) -- node[anchor=south] { } (yic);
\draw [->, gray] (xl) -- node[anchor=south] { } (yl);
\draw [->, gray] (mu) -- node[anchor=south] { } (xh);
\draw [->, gray] (xh) -- node[anchor=south] { } (xic);
\draw [->, gray] (xic) -- node[anchor=south] { } (xl);
\node (nc)[namecompartment, below of=PL, yshift=-1.3cm, xshift=1.5cm] {\tiny $t=2, \dots, T$};
\end{tikzpicture}
\caption{\Ac{DAG} representation of a stochastic-severity model.
}
\label{fig3b}
\end{subfigure}\\
	\begin{subfigure}[c]{11cm}
	\centering
	\begin{tikzpicture}[node distance=1.8cm]
\node (s)[comp,  draw opacity=0.4]{\color{gray}$S$};
\node (e)[comp, draw opacity=0.4,  right of=s, xshift=0.1cm] {\color{gray}$E$};
\node (i)[comp,  draw opacity=0.4, right of=e, xshift=.1cm] {\color{gray}$I$};
\node (r)[comp,  draw opacity=0.4, right of=i, xshift=.1cm] {\color{gray}$R$};
\draw [gray, ->, thick,] (s) -- node[anchor=south] {$x^\textsc{v}$} (e);
\draw [gray, ->, thick] (e) -- node[anchor=south] { } (i);
\draw [gray, ->, thick] (i) -- node[anchor=south] { } (r);
\node (h)[comp, below of=e, , xshift=0.5cm, yshift=.5cm] {$ \textsc{h} $};
\node (ic)[comp, right of=h] {$ \textsc{ic} $};
\node (d)[comp, right of=ic] {$ \textsc{d} $};
\draw[ ->, thick] (e.west) .. controls +(down:0mm) and +(left:20mm) .. (h.west);
\draw [black, ->, thick,] (h) -- node[anchor=south] { } (ic);
\draw [black, ->, thick,] (ic) -- node[anchor=south] { } (d);
	\end{tikzpicture}
	\caption{\rev{Compartmental model representation of a stochastic severity model: individuals entering the Exposed state can also start the severity pathways (bottom), in parallel to their transmission pathway (top) }
	}
	\label{fig3c}
\end{subfigure}
\caption{\rev{Pyramid (a), \ac{DAG} (b) and compartmental model (c) representations of a severity process including viral infection ($ \textsc{v} $), hospitalisation {($ \textsc{h} $)}, intensive care admission {($ \textsc{ic} $)}, and death {($ \textsc{d} $)}.}}
\label{f3}
\end{figure}

\rev{A simple model, such as the one presented in Figure \ref{fig3b}, assumes that the number of people in a severity category is a subsample (e.g. a Binomial sample) of the number of people in the preceding less-severe category with severity parameters $ \rho^{\textsc{h}|\textsc{v}}, \rho^{\textsc{ic}|\textsc{h}}, $ and $\rho^{\textsc{d}|\textsc{ic}}$ representing the risk of moving from one severity state to the next one.
The observational processes describes the number of detected cases $ Y^\textsc{h}_{1:t} $, $ Y^\textsc{ic}_{1:t} $, and $ Y^\textsc{d}_{1:t} $ in each specific severity layer as a subsample of the cases in the previous layer with a layer-specific detection probability $ \zeta^\textsc{h} $, $ \zeta^\textsc{ic} $, $ \zeta^\textsc{d} $. In Figure \ref{fig3b}} the connection between multiple data can be easily spotted: i.e. not only are they linked through the underlying infection process, but also via the progression through severity states. In fact, while the likelihood of each of multiple datasets given the infection process $ X^\textsc{v}_{1:t} $ is typically available, these data are not independent, conditionally on the parameters: they share common severity processes \rev{$ X^\textsc{h},X^\textsc{ic}, X^\textsc{d} $}, and therefore their joint likelihood cannot be simply derived via the product of each likelihood component.

\subsubsection{Dependency between data streams across time}
Another challenge is given by delays between events: severe events usually do not happen simultaneously. More likely, there will be some time elapsing between events of increasing severity; this could lead to, for example, cases detected \rev{upon their hospitalisation at time $ t $ also being present in data on \ac{IC} admission at some time $ t+s , s\geq 0$ }. This introduces \rev{additional dependence between observations across time. }

\rev{To see this, consider the example of hospitalisation and \ac{IC} admissions as events of increasing severity occurring in succession. Let $X^{\textsc{ic}|\textsc{h}_t} $ be the number of people that will eventually experience \ac{IC} admission ($ \textsc{ic} $), having already experienced hospitalisation at $ t $ ; and let $ X^{\textsc{ic}|\textsc{h}_t}_s $ be the subset experiencing \ac{IC} admission at time $ s\geq t $. 

The number of people $ X^\textsc{ic}_s $ being admitted to \ac{IC} at $ s $ irrespectively of the time of hospitalisation, is $ X^{\textsc{ic}}_s = \sum_{t\leq s} X^{\textsc{ic}|\textsc{h}_t}_s  $. 

These \acp{r.v.} can be assumed to be generated according to the following model: starting from the number of people $ X^\textsc{h}_t $, that have had event $ \textsc{H} $ at time $ t $ ($ t=1,\dots, T $), the number $X^{\textsc{ic}|\textsc{h}_t}$ that will experience event $ \textsc{ic} $, irrespectively of its time, is distributed according to a Binomial \ac{r.v.} with the corresponding severity probability $ \rho^{\textsc{ic}|\textsc{h}} $:}
\begin{equation}\label{eq9}
X^{\textsc{ic}|\textsc{h}_t}|x^\textsc{h}_t \sim \text{Bin}(x^{\textsc{h}}_t,   \rho^{\textsc{ic}|\textsc{h}} ) 
\end{equation}
for $ t=1, \dots, T $. 

To describe the time elapsing between the $ \textsc{h} $ and $\textsc{ic}$ events, denote by $ f^{\textsc{ic}|\textsc{h}} _d(\boldsymbol{\varphi}^{\textsc{ic}|\textsc{h}} ) $ the probability that the delay between event is in the $ d $th interval of length $ \delta $, $ [\delta d; \delta d+\delta) $ for $ d=0,1,\dots, D $, with $ D $ being the largest interval index for which the delay is relevant (i.e. $ f^{\textsc{ic}|\textsc{h}} _d(\boldsymbol{\varphi}^{\textsc{ic}|\textsc{h}} )\approx0 $ for $ d>D $). $ f^{\textsc{ic}|\textsc{h}} _d(\boldsymbol{\varphi}^{\textsc{ic}|\textsc{h}} ) $ is often derived from the discretization of a parametric distribution with appropriate parameter vector $ \boldsymbol{\varphi}^{\textsc{ic}|\textsc{h}}  $. To simplify notation, $\boldsymbol{\varphi}^{\textsc{ic}|\textsc{h}}  $ is dropped, with $ f^{\textsc{ic}|\textsc{h}} _d $ representing both the function and the parameters used to describe the delay from hospitalisation to \textsc{IC} admission. 

\rev{Under this discrete definition of the distribution of the time to event, and conditionally on $ X^{\textsc{ic}|\textsc{h}_t}$, the introduction of stochastic delays could be allowed by defining 
$ X^{\textsc{ic}|\textsc{h}_t}_s $, as a component of the Multinomial \ac{r.v.}: 
\begin{equation}\label{eq10}
\begin{bmatrix}
	X^{\textsc{ic}_{t}|\textsc{h}_t}\\
	X^{\textsc{ic}_{t+1}|\textsc{h}_t}\\
	\dots\\
	X^{\textsc{ic}_{t+D}|\textsc{h}_t}
\end{bmatrix}\bigg| x^{\textsc{ic}|\textsc{h}_t} \sim \text{Multi} \left(x^{\textsc{ic}|\textsc{h}_t},
\begin{bmatrix}
	f^{\textsc{ic}|\textsc{h}}_{0} \\
	f^{\textsc{ic}|\textsc{h}}_{1}\\
	\dots\\
	f^{\textsc{ic}|\textsc{h}}_{D}
\end{bmatrix} \right)\\
\end{equation}
with $ s=t+d $. The number of people $X^\textsc{ic}_{t}$ that are admitted to \ac{IC} at each time $ t=1, \dots T $ can then be obtained by summing these stochastic terms, i.e.:
\begin{equation}\label{eq11}
X^\textsc{ic}_{t}= \sum_{d=0}^{D} X^{\textsc{ic}|\textsc{h}_{t-d}}_{t}
\end{equation}}
which can be recognised as a typical stochastic convolution to describe delays in epidemic models \citep{brookmeyer1994aids}. The counts of events at a higher level of severity can be modelled likewise.

\rev{From Equation \eqref{eq10} the time dependence between events is evident: here} the number of people experiencing $ \textsc{ic} $ at time $ t $ depends on \acp{r.v.} defined on the previous $ D $ intervals. 

\subsection{A model for multiple data on severe influenza} 
Influenza is monitored by many surveillance schemes in the UK, one of these is the Severe Acute Respiratory Infection (SARI)-Watch scheme \citep{SARIWatch}, which has evolved from the \ac{USISS}, the main source of data on severe influenza cases in the UK prior to the COVID-19 pandemic. According to the \ac{USISS} protocol \citep{health2011sourcesa}, all \ac{NHS} trusts in England report  the weekly number of laboratory-confirmed influenza cases admitted to \ac{IC} units. In addition to this mandatory scheme, a sentinel subgroup of NHS trusts in England is recruited every year to participate in a sentinel scheme \citep{health2011sourcesb, boddington2017developing}, which reports weekly numbers of laboratory-confirmed influenza cases hospitalised at all levels of care. Some individuals might be detected in both datasets, leading to a dependence \rev{(see Figure 1 in the SI)}. 

\subsubsection{Parameterisation and data generation}Denote by $ \vtheta $ the set of parameters, composed of $ \vtheta^T$, the parameters of a $SEIR$ transmission model, tracking the number of susceptible ($S$), exposed ($E$), infected ($I$) and removed ($R$), individuals, and $ \vtheta^S $, the parameters of the severity and detection model. \rev{The transmission and severity models are here considered separately with individuals potentially entering the severity pathway upon infection and then proceeding to have events of increasing severity while still progressing across the $ SEIR $ states (see Figure \ref{fig3c} above).}

$ \vtheta^T=\left\lbrace \beta , \sigma, \gamma ,\pi, \iota\right\rbrace   $  consists of the transmission rate $ \beta $; the exit rates from compartments $ E $ and $ I $, $ \sigma $ and $ \gamma $ respectively; and the initial proportions of immune, $\pi$, and of infected/infectious individuals, $ \iota $. The parameters $\pi$ and $ \iota $, together with $ \sigma $, $ \gamma $ and the known constant $ N $, the total size of the population, contribute to the formulation of the initial state of the epidemic. Seasonal influenza is a large epidemic that takes place every winter with high likelihood, hence a deterministic transmission model is assumed: the information contained in $ \vtheta^T $, together with the known constants, provides the full time series of new infections. Let the time be discretised in intervals of length $ \delta $ so that the $ t $-th interval covers the time $ [t\delta , t\delta+\delta) $ and the intervals are indexed by $ t=0,1,2, \dots, T $, where $ t=0 $ coincides with the beginning of the data collection period and $ t=T $ is the end of the data collection period. Denote the number of susceptible individuals at the beginning of interval $ t $ by $ S_t $ and likewise for the other compartments $E,I,R $. Denote by $ \xi^\textsc{v}_{1:T} $ the vector of the number of new infections in interval $ t=0,1,2,\dots T $. 

$ \vtheta^S=\left\lbrace \rho^{\textsc{h}|\textsc{v}}, \rho^{\textsc{ic}|\textsc{h}}, f^{\textsc{h}|\textsc{v}}, f^{\textsc{ic}|\textsc{h}}, \zeta_t^\textsc{h}, \zeta_t^\textsc{ic} \right\rbrace  $ includes severity and detection parameters: $ \rho^{\textsc{h}|\textsc{v}}$ is the probability of hospitalization given viral infection; $ \rho^{\textsc{ic}|\textsc{h}} $ is the probability of \ac{IC} admission given hospitalization; $ f^{\textsc{h}|\textsc{v}}$ and $ f^{\textsc{ic}|\textsc{h}} $ denote probabilities for the times from infection to hospitalisation and from hospital admission to \ac{IC} admission, respectively; $ \zeta_t^\textsc{h} $ and $ \zeta_t^\textsc{ic} $ are the probability of detecting an hospitalised and \ac{IC} case, respectively.

\subsubsection{\ac{SSM} formulation}
Describing the model using \ac{SSM} notation, the state process is composed of the distributions of: $  X^{\textsc{h}|\textsc{v}_t}$ , the number of hospitalizations that had a viral infection at each interval $ t $; $ X^\textsc{h}_{t} $, the number of hospitalizations at $ t $, obtained via a convolution of $ X^{\textsc{h}|\textsc{v}_t}$; the number of hospitalizations that eventually will be admitted to IC and have been hospitalised at $ t $, $ X^{\textsc{ic}|\textsc{h}_t} $; the number of IC admissions at $ t $, $ X^\textsc{ic}_t$ obtained by the convolution of  $X^{\textsc{ic}|\textsc{h}_t} $, for $ t=0,1,2,\dots, T $. This state process is assumed to be:
\begin{equation}\label{eq12}
\begin{split}
		X^{\textsc{h}|\textsc{v}_t}&\sim \text{Pois}(\xi^{\textsc{v}}_t\cdot  \rho^{\textsc{h}|\textsc{v}} )\\
		X^\textsc{h}_{t}& = \sum_{s=0}^{S} {X_{t}^{\textsc{h}|\textsc{v}_{t-s}}}\qquad
		\text{with }	\begin{bmatrix}
			X^{\textsc{h}_{t}|\textsc{v}_t}\\
			X^{\textsc{h}_{t+1}|\textsc{v}_t}\\
			\vdots\\
			X^{\textsc{h}_{t+D}|\textsc{v}_t}
		\end{bmatrix}\bigg|   x^{\textsc{h}|\textsc{v}_t} \sim \text{Multi} \left(x^{\textsc{h}|\textsc{v}_t},
		\begin{bmatrix}
			f^{\textsc{h}|\textsc{v}}_{0} \\
			f^{\textsc{h}|\textsc{v}}_{1}\\
			\vdots\\
			f^{\textsc{h}|\textsc{v}}_{D}
		\end{bmatrix} \right)\\
		X^{\textsc{ic}|\textsc{h}_t}&\sim \text{Bin}(X^{\textsc{h}}_t,  \rho^{\textsc{ic}|\textsc{h}} )\\
		X^\textsc{ic}_{t}& = \sum_{s=0}^{S} {X_{t}^{\textsc{ic}|\textsc{h}_{t-s}}}\qquad
		\text{with }\begin{bmatrix}
			X^{\textsc{ic}_{t}|\textsc{h}_t}\\
			X^{\textsc{ic}_{t+1}|\textsc{h}_t}\\
			\vdots\\
			X^{\textsc{ic}_{t+D}|\textsc{h}_t}
		\end{bmatrix}\bigg| x^{\textsc{ic}|\textsc{h}_t} \sim \text{Multi} \left(x^{\textsc{ic}|\textsc{h}_t},
		\begin{bmatrix}
			f^{\textsc{ic}|\textsc{h}}_{0} \\
			f^{\textsc{ic}|\textsc{h}}_{1}\\
			\vdots\\
			f^{\textsc{ic}|\textsc{h}}_{D}
		\end{bmatrix} \right)
\end{split}
\end{equation}
for $ t=0,1,\dots, T $, \rev{where the first line is given by the approximation of a Binomial sample, similar to the one in Equation \eqref{eq9}, to the respective Poisson distribution. }Here $ 	f^{\textsc{h}|\textsc{v}}_{0:D} $ and $ 	f^{\textsc{ic}|\textsc{h}}_{0:D} $ are vectors containing elements $ 	f^{\textsc{h}|\textsc{v}}_{d} $ and $ 	f^{\textsc{ic}|\textsc{h}}_{d} $ denoting the probability of experiencing a delay of $ d $ weeks between infection and hospitalization and hospitalization and \ac{IC} admission, respectively, for $ d=0, \dots, D $. These are considered known and fixed. 

The observational process consists of the distributions of two datasets, $ y_{1:T}^\textsc{h} $, the count of hospitalizations, and $ y_{1:T}^\textsc{ic}  $, the count of IC admissions, conditional on the hidden states $ X_{1:T}^\textsc{h} $ and $ X_{1:T}^\textsc{ic}  $. These are  assumed to be Binomial with detection probabilities $ \zeta^\textsc{h}_t $ and $ \zeta^\textsc{ic}_t $ respectively:
\begin{equation}\begin{split}\label{eq13}
Y_{t}^\textsc{h}| x^\textsc{h}_{t}  &\sim \text{Bin} (x^\textsc{h}_{t}, \zeta^\textsc{h}_t)\\
Y_{t}^\textsc{ic}|X x^\textsc{ic}_{t}&\sim \text{Bin} (x^\textsc{ic}_{t}, \zeta^\textsc{ic}_t)
\end{split}
\end{equation}
for $ t=0,1, 2, \dots, T$\rev{. This implies that individuals enter a dataset upon reaching a specific severity state (e.g. their \ac{IC} admission is registered on the day/week when they entered \ac{IC})}. \rev{Here $ \zeta^\textsc{h}_t $ is proportional to the probability of ending up in a sentinel hospital, and therefore be registered in the sentinel data scheme. } 

\subsection{Inference} Two inferential methods are proposed here: the first does not account for the dependence among data while the second aknowledges this dependence properly encapsulating the layers of stochasticity present in the model. 

\subsubsection{The approximation under an independence assumption} 
\rev{With data generated by such a scheme as above, introducing the erroneous assumption that data on hospitalizations and \ac{IC} are independent, the likelihood might be available in closed form, which would substantially simplify inference. }
Thanks to the properties of the Poisson process \citep{kingman1992poisson}, several hidden states and the data distribute marginally according to a Poisson distribution:
\begin{equation}\begin{split}
X^\textsc{h}_{t}&\sim \text{Pois} \left( \rho^{\textsc{h}|\textsc{v}} \cdot\sum_{d=0}^{D}  \xi^\textsc{v}_{t-d} \cdot f^{\textsc{h}|\textsc{v}}_{d} \right) \\
X^\textsc{ic}_{t} &\sim \text{Pois}\left(\rho^{\textsc{ic}|\textsc{h}}\cdot\rho^{\textsc{h}|\textsc{v}} \cdot\sum_{d=0}^{D} \sum_{g=0}^{d}  \xi^\textsc{v}_{t-d-g} \cdot f^{\textsc{h}|\textsc{v}}_{d}\cdot f^{\textsc{ic}|\textsc{h}}_{g}\right) \\
Y_{t}^\textsc{h} &\sim \text{Pois} \left( \zeta^\textsc{h}_t\cdot\rho^{\textsc{h}|\textsc{v}} \cdot\sum_{d=0}^{D}  \xi^\textsc{v}_{t-d} \cdot f^{\textsc{h}|\textsc{v}}_{d} \right) \\
Y_{t}^\textsc{ic} &\sim \text{Pois}\left(\zeta^\textsc{ic}_t\cdot \rho^{\textsc{ic}|\textsc{h}}\cdot \rho^{\textsc{h}|\textsc{v}} \cdot\sum_{d=0}^{D} \sum_{g=0}^{d}  \xi^\textsc{v}_{t-d-g} \cdot f^{\textsc{h}|\textsc{v}}_\textsc{d}\cdot f^{\textsc{ic}|\textsc{h}}_{g}\right)
\end{split}
\label{eq14}
\end{equation}
for $ t=0,1, 2, \dots T $.

Ignoring the dependence between the two data streams, the joint likelihood is
\begin{equation}
p\left(y_{1:T}^\textsc{h}, y_{1:T}^\textsc{ic} \bigg| \boldsymbol{\theta}^T , \boldsymbol{\theta}^S \right)= p\left(y_{1:T}^\textsc{h} \bigg| \boldsymbol{\theta}^T , \boldsymbol{\theta}^S  \right)\times p\left(y_{1:T}^\textsc{ic} \bigg| \boldsymbol{\theta}^T , \boldsymbol{\theta}^S  \right),
\label{eq15}
\end{equation}
where each factor is the Poisson density of Equation \eqref{eq14}. 

\rev{The independence assumption allows more rapid inference, with a likelihood available in closed form; 
in of Section \ref{sec4} we assess if this approximation induces errors in terms of bias or uncertainty quantification. }
\subsubsection{Exact inference via pseudo-marginal methods} To properly account for dependence and stochasticity, a simulation algorithm is proposed to approximate the joint likelihood of the hospitalization and IC data. 
Their joint probability distribution can be decomposed in two ways:
\begin{equation*}
\begin{split}
p\left(y_{1:T}^\textsc{h}, y_{1:T}^\textsc{ic}\bigg|  \boldsymbol{\theta}^T , \boldsymbol{\theta}^S\right)&=p\left(y_{1:T}^\textsc{h}\bigg|  y_{1:T}^\textsc{ic}, \boldsymbol{\theta}^T , \boldsymbol{\theta}^S\right)\times p\left( y_{1:T}^\textsc{ic}\bigg|  \boldsymbol{\theta}^T , \boldsymbol{\theta}^S\right)\\
&=p\left(y_{1:T}^\textsc{ic}\bigg|  y_{1:T}^\textsc{h}, \boldsymbol{\theta}^T , \boldsymbol{\theta}^S\right)p\left( y_{1:T}^\textsc{h}\bigg|  \boldsymbol{\theta}^T , \boldsymbol{\theta}^S\right)
\end{split}
\end{equation*}
where, in both cases, the second of the two factors is available in closed form (Equation \ref{eq14}).

To approach the estimation of the other factor, state-inference methods for \acp{SSM} can be used. The methods for inference described in Section \ref{sec2} address the Markovian dependence across time. Here however, the time-dependence of the transmission process disappears thanks to the deterministic approximation. However, the dependence over the severity domain remains: the distributional assumptions of Equation \eqref{eq12} can be used to construct a simulation-based estimator of the likelihood that sequentially approximates severity states.

\rev{More specifically, given that the marginal distribution $ p(y_{1:T}^\textsc{ic}|  \vtheta) $ is available in closed form, an estimate of the conditional distribution $ p(y_{1:T}^\textsc{h}|  y_{1:T}^\textsc{ic}, \vtheta) $ can be obtained  sampling from the hidden states . This is done via the following steps: \textit{(i)} sample $ N $ particles from the distribution of the \ac{IC} admissions conditionally on the data $ y_{1:T}^\textsc{ic}$ (this can be done directly as shown in the Supplementary Information, sec.1 as the unobserved number of \ac{IC} admissions is also Poisson distributed); \textit{(ii)} These particles can then be back projected to their date of hospitalization, using the discrete delay distributions introduced in Equation \eqref{eq12}; therefore we can\textit{ (iii)} obtain a sample from the distribution of the number of people who are hospitalised at time $ t $ and will be admitted to \ac{IC} in the future, conditionally on \ac{IC} data; from those samples we can finally \textit{(iv) }sample the time series of hospitalizations $ x_{1:T}^\textsc{h} $, by simulating all those individuals that were admitted to hospital but not to \ac{IC} (using the same properties of step\textit{ (i)}). Algorithm \ref{alg2} reports a pseudo-code of this scheme, where all the steps are explicitly mentioned. 
}

\begin{algorithm}[h]
\SetAlgoLined
\KwResult{$ \widehat{p}(y_{1:T}^\textsc{h},  y_{1:T}^\textsc{ic}| \vtheta) $}
\KwIn{fixed parameter $\vtheta $, number of particels $ {N} $, data $ y_{1:T}^\textsc{h} $,  $ y_{1:T}^\textsc{ic} $}
compute\\
$ p(y_{1:T}^\textsc{ic}| \vtheta) $ =  $ f(y_{1:T}^\textsc{ic}|\zeta^\textsc{ic}_t\cdot \rho^{\textsc{ic}|\textsc{h}}\cdot \rho^{\textsc{h}|\textsc{v}}\cdot \sum_{d\text{=}0}^{D} \sum_{g\text{=}0}^{d}  \xi^\textsc{v}_{t\text{-}d\text{-}g} \cdot f^{\textsc{h}|\textsc{v}}_{d}\cdot f^{\textsc{ic}|\textsc{h}}_{g} ) $ with $ f(\cdot) $ being a Poisson density\\
\For{$ n=1,\dots, {N} $}{
\For{$ t=0, 1, \dots T $}{\vspace*{0.2cm}
sample : $ {x^\textsc{ic}_{t}}^{(n)}  \sim$ Pois $ \left(\left(1-\zeta^\textsc{ic}_t\right) [\rho^{\textsc{ic}|\textsc{h}}\cdot \rho^{\textsc{h}|\textsc{v}} \sum_{d=0}^{D} \sum_{g=0}^{d}  \xi^\textsc{v}_{t-d-g} \cdot f^{\textsc{h}|\textsc{v}}_{d}\cdot f^{\textsc{ic}|\textsc{h}}_{g} ]  \right) +y_{T}^\textsc{ic} $ \\\vspace*{0.2cm}
sample : $ {x^{\textsc{ic}_{t}|\textsc{h}_{t}}}^{(n)}\text{, }{x^{\textsc{ic}_{t}|\textsc{h}_{t-1}}}^{(n)} \dots\text{, } {x^{\textsc{ic}_{t}|\textsc{h}_{t-D}}}^{(n)} \sim \text{Multi} \left({x^\textsc{ic}_{t}}^{(n)},  f^{\textsc{ic}|\textsc{h}}_{0:D} \right) $\\\vspace*{0.2cm}
compute : 
${ x^{\textsc{ic}|\textsc{h}_t}}^{(n)}=\sum_{d=0}^{D} {x^{\textsc{ic}_{t+d}|\textsc{h}_{t}}}^{(n)} $\\
\vspace*{0.2cm}
sample : 
$ {x^\textsc{h}_{t}}^{(n)}| {x^{\textsc{ic}|\textsc{h}_t}}^{(n)}\sim \text{Pois}\left(\left(1-\rho^{\textsc{ic}|\textsc{h}} \right)\left[ \rho^{\textsc{h}|\textsc{v}}\sum_{d=0}^{D}  \xi^\textsc{v}_{t-d} \cdot f^{\textsc{h}|\textsc{v}}_{d} \right] \right) +{x^{\textsc{ic}|\textsc{h}_t}}^{(n)} $
}compute : $ p\left(y_{1:T}^\textsc{h}|{x^\textsc{h}_{1:T}}^{(n)}\right) $ = $g\left(
y_{1:T}^\textsc{h}|{x^\textsc{h}_{1:T}}^{(n)}, \zeta^\textsc{h}_t\right) $
with $ g(\cdot) $ being a Binomial density
}
$ \widehat{p}\left(y_{1:T}^\textsc{h}, y_{1:T}^\textsc{ic}|  \vtheta\right)= p\left(y_{1:T}^\textsc{ic}| \vtheta\right)\cdot \frac{1}{N}\sum_{n=1}^{{N}}p\left(y_{1:T}^\textsc{h}|{x^\textsc{h}_{1:T}}^{(n)}, \vtheta\right)$
\caption{Approximation of the likelihood ${p}(y_{1:T}^\textsc{h}, y_{1:T}^\textsc{ic}|  \vtheta)$}
\label{alg2}
\end{algorithm}

\rev{A second algorithm uses the alternative factorization of the joint likelihood. While still feasible, this algorithm performs more poorly than Algorithm \ref{alg2}, mainly because the \ac{IC} admissions are more-completely observed and smaller in numbers, hence simulation methods tend to fail \citep{brooks2011handbook}. In contrast, Algorithm \ref{alg2}, provides a stable approximation of the joint distribution and has been observed empirically to perform well in simulated and real-data scenarios. }
See the Supplementary information for the derivation of the two algorithms.

\section{Relevance of the dependence}\label{sec4}
To assess whether (and in which situations) accounting for the dependence makes any difference {compared to} assuming the two datasets to be independent a full simulation study is set up t. Intuitively, a miss-specified model that does not account for dependencies would assume more independent information than is truly present in the data, hence would achieve overly-confident results. Conversely, truly accounting for dependence should reflect more properly the unknowns and noise of the model and also help the estimation of those parameters that effectively link the dependent data streams. 

\subsection{Simulation study set-up}
Data are simulated to reflect a situation similar to the motivating \ac{USISS} data on a smaller population, chosen  to reduce the computation time. The datasets are generated with some common parameters and some scenario-specific parameters. 

The common parameters are:
\begin{equation*}
\left\lbrace N=10000,\beta=0.63, \pi=0.3, \iota=0.0001, \sigma=\frac{1}{4}, \gamma=\frac{1}{3.5}, f^{\textsc{h}|\textsc{v}} \sim \text{Exp}(0.3), f^{\textsc{ic}|\textsc{h}}\sim \text{Exp}(0.4) \right\rbrace 
\end{equation*}
while the scenario specific parameters are reported in Table \ref{t1}. Each of the severity and detection parameters can take either a small or a large value. The smaller leads to situations where the probability of being observed in both datasets is low, therefore data are less dependent; while when parameters take larger values, there is more overlap between datasets and therefore more dependence.
\begin{table}[h]
\caption{Parameters used to generate the datasets.}
\label{t1}
\centering
\begin{tabular}{c c c}
&small dependence&large dependence\\
\hline
$\rho^{\textsc{h}|\textsc{v}}$&0.1&0.5\\
$\rho^{\textsc{ic}|\textsc{h}}$&0.1&0.9\\
$ \zeta_t^\textsc{h}=\zeta^\textsc{h}$&0.1&0.3\\
$\zeta_t^\textsc{ic}=\zeta^\textsc{ic}$&0.1&0.9\\
\hline
\end{tabular}
\end{table}
The values have been chosen to be at the extremes of a realistic range: e.g., the probability of hospitalization and the detection of hospitalization in the {large dependence} case are smaller than the respective quantities for \ac{IC} unit admission because usually more severe cases are better monitored, and hence have a higher detection.

The aim of the comparison is to assess whether a misspecified independent likelihood (Equation \ref{eq13}) would affect the inference of the parameters, leading to different posterior distributions from the ones obtained with the \ac{MC} approximation of the joint likelihood assuming  dependent data (Algorithm \ref{alg2}). For this reason the results presented here are to be compared {within} each scenario: between the ones obtained with the independent misspecified model (abbreviated with \textsc{miss ind}) and the ones obtained using the dependent joint model (abbreviated with \textsc{joint dep}). 

\subsection{Results of the simulation study} As outlined in detail below, the misspecified model leads to overly precise results in the estimation of the transmission parameters when the dependence is large. Furthermore, the parameter connecting the two severity states to which the data refer is estimated with less precision when the misspecified model is assumed.

\subsubsection{Comparison for transmission parameters}
For the parameter inference with the approximated joint likelihood, a \ac{MCWM} algorithm with likelihood approximation via Algorithm \ref{alg2} with $ N=2000 $ was chosen. 
Resulting estimates were compared with those from the misspecified independent Poisson likelihood over 1000 synthetic datasets. 500 datasets were simulated using the smaller values of the parameters (left column of Table \ref{t1}) and 500 datasets using the larger values (right column of Table \ref{t1}). The only parameters inferred are the transmission parameters $ \beta, \iota $ and $ \pi $, with the severity parameters being fixed at their true, scenario-specific, value.

In the case of {small dependence}, the results show that the posterior distributions obtained with the misspecified independent likelihood are very similar to the ones obtained with the approximated joint likelihood. To measure any discrepancy between the two estimation methods, the pairwise difference (PWD) in variance Var$ (\hat{\alpha}^{m} \mid y^d) $ for $ m \in \{\textsc{joint dep}, \textsc{miss ind}\} $ and $ \alpha $ being one of the parameters, and in the width R${}_{95}(\hat{\alpha}^{m} \mid y^d) $ of the 95\% \ac{CrI} was computed as:
\begin{equation*}\begin{split}
\text{PWD}(\text{Var}(\alpha))^d&= \text{Var}(\widehat{\alpha}^\textsc{joint dep}|\boldsymbol{y}^d)-\text{Var}(\widehat{\alpha}^\textsc{miss ind}|\boldsymbol{y}^d) \qquad \alpha= \beta, \pi, \dots ; d=1,2,\dots, 500 \\
\text{PWD}(\text{R}_{95}(\alpha))^d&= \text{R}_{95}(\widehat{\alpha}^\textsc{joint dep}|\boldsymbol{y}^d)-\text{R}_{95}(\widehat{\alpha}^\textsc{miss ind}|\boldsymbol{y}^d) \qquad \alpha= \beta, \pi, \dots ; d=1,2,\dots, 500.
\end{split}
\end{equation*} 

\rev{If the assumption of independence between data does not affect the inference, then the variances and \ac{CrI} range would be equal and therefore PWD would be 0. If instead the \textsc{miss ind} routine leads to error, we would expect the posterior variability measures to be lower compared to the exact inference obtained by the \textsc{joint dep}, and therefore PWD would be positive.}

\begin{figure}[h]
\hspace*{-1.5cm}\includegraphics[scale=0.32,page=4]{ss_sec4.pdf}\includegraphics[scale=0.32, page=10]{ss_sec4.pdf}\\
\hspace*{-1.5cm}\includegraphics[scale=0.32,page=1]{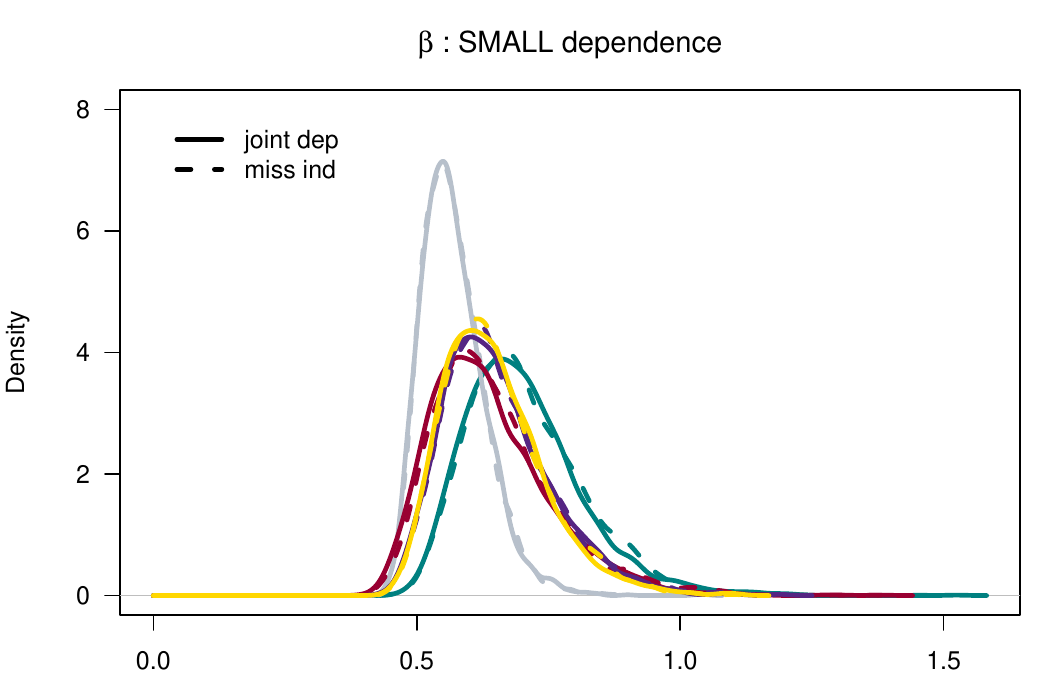}\includegraphics[scale=0.32, page=7]{ss_sec4.pdf}
\caption{Histogram of the pairwise differences in variance and in 95\% \ac{CrI} length of the posterior distribution of the transmission parameters $ \beta $ for the small dependence scenario (left) and big dependence scenario (right). The bottom plots report the posterior distribution estimated with the \textsc{joint dep} (solid) and \textsc{miss ind} (dashed) model for 5 randomly-selected simulated data (colors).}
\label{f5}
\end{figure}

These quantities are reported in Figure \ref{f5} (top) which shows an imperceptible difference in the posterior distributions and their precision-summaries between the two models. 

The same analysis is run on the 500 datasets with a {large dependence}, with results reported in Figure \ref{f5} (bottom). 
Here there is a notable difference between the results from the two models: the posterior distributions from the misspecified model that assumes independent data are less variable than the ones derived using the \ac{MC} approximation of the joint dependent likelihood. 

This result was expected, since the misspecified model, by assuming independent data, accounts for more information than is contained in the data. This leads to an overconfidence that can be detected in the underestimation of the posterior variance. Results are confirmed by the proportion of datasets for which the pairwise differences are less than or equal to 0 for each of the parameters (Table \ref{t2}). When this quantity is close to 0.5, the variances of the estimates obtained with the two methods are similar within datasets; when this quantity is close to 1 it suggests that the variance of the estimates obtained using the misspecified independent likelihood is systematically larger than the variance of the estimates obtained with the joint likelihood; and when this quantity is close to 0 it highlights that the former variance is systematically smaller then the latter. 
\begin{table}[h]
\caption{Proportion of datasets in which the pairwise difference of variance is smaller or equal to 0 for the three transmission parameters.} 
\label{t2}
\centering
\begin{tabular}{c c c}
{Parameter} & \multicolumn{2}{c}{Proportion of ${\textsc{pwd}(\text{Var})}\leq 0 $} \\
\hline
&{Small dependence}&{Large dependence}\\
\hline
$ {\beta}$&0.392&0 \\
$ {\pi}$&0.390&0 \\
$ {\iota} $&0.378&0 \\
\hline
\end{tabular}
\end{table}
The results strongly suggest a systematic difference in variability between the two methods in the large dependence scenario. In all the simulated datasets the variances of the posterior distributions of the parameters are smaller in the analysis using the misspecified independent model than in the approximate joint model.

\subsubsection{Comparison for transmission and severity parameters}
The same comparison is carried out in a context where inference is drawn both for the transmission and the severity parameters. Here, since more quantities are estimated and due to the high correlation of the parameters of epidemic models, a difference between the results from the two models may be more difficult to spot. Moreover, in this multi-parameter context, convergence is sometimes compromised, particularly in the large-dependence scenario. 
In the distribution of the pairwise differences, neither for the transmission parameters nor for the newly estimated severity parameters, can a large difference be seen (figures reported in the Supplementary Information). 

For the large dependence scenario, the only notable difference can be seen in the distribution of $ \rho^{\textsc{ic}|\textsc{h}}$: the parameter that links the two datasets, since it defines the probability of \ac{IC} admission conditional on hospitalization. When the two datasets are jointly analysed, they both contribute to the estimation of $ \rho^{\textsc{ic}|\textsc{h}}$, with hospital data informing the Binomial size in Equation \eqref{eq13} and \ac{IC} data informing the proportion of people in the more-severe state. When the two datasets are considered independently, the hospital data do not play any role in the inference of $ \rho^{\textsc{ic}|\textsc{h}}$.The proportions of pairwise differences less than or equal to 0 confirm this: the variance of the posterior sample of the parameter $\rho^{\textsc{ic}|\textsc{h}}$ is always lower when inference is drawn with the approximation to the joint dependent likelihood compared to when the misspecified independent model is adopted {(Table \ref{t3})}. 

\begin{table}[!h]
\caption{Proportion of datasets in which the pairwise difference of variance is smaller or equal to 0 for the transmission and severity parameters. }
\label{t3}
\centering
\begin{tabular}{c c c} 
{Parameter} &  \multicolumn{2}{c}{Proportion of $ {\textsc{pwd}(\text{Var})}\leq 0 $ }\\
\hline
&{Small dependence}&{Large dependence}\\
\hline
$ {\beta}$&0.498& 0.554\\
$ {\pi}$&0.464& 0.546\\
$ {\iota} $&0.348& 0.202\\
$ \rho^{\textsc{h}|\textsc{v}} $&0.466&0.566\\
$ \rho^{\textsc{ic}|\textsc{h}} $&0.778& 1\\
\hline
\end{tabular}
\end{table}

\subsubsection{Influential parameters}
As a final comparison, a further investigation into the main cause of the difference is {undertaken}. Starting from the small-dependence scenario, one at a time, each parameter of Table \ref{t1} is allowed to take the larger value in simulating the 500 datasets.

Estimates of the {five} parameters are then obtained according to the misspecified independent and the joint dependent model. The posterior distributions and the plots of the precision statistics are reported in the Supplementary Information. While a detectable difference in the results is observed when all the parameters affecting the level of dependence vary, the same cannot be said when each parameter increases alone. Differences are less evident, with the probability of detection in \ac{IC} being the most influential parameter, as shown in {Table \ref{t4}}, where each column corresponds to a scenario where all the parameters but the header of the column are assumed small. 

\begin{table}[!h]
\caption{Proportion of datasets in which the pairwise difference of variance is smaller or equal to 0 for the transmission and severity parameters in the scenario with small dependence except for the respective column-name parameter.}
\label{t4}
\centering
\begin{tabular}{c c c c c}
Increased Parameter &\hspace*{.5cm} $ \rho^{\textsc{h}|\textsc{v}} $ \hspace*{.5cm}&\hspace*{.5cm}$ \rho^{\textsc{ic}|\textsc{h}}$\hspace*{.5cm} &\hspace*{.5cm} $ \zeta^\textsc{h} $\hspace*{.5cm}&\hspace*{.5cm}$ \zeta^\textsc{ic} $\\
\hline
{Parameter} &\multicolumn{4}{c}{Proportion of $ {\textsc{pwd}(\text{Var})}\leq 0 $ }\\
\hline
$ \beta $&0.468   &0.454 &0.296 & 0.490 \\
$ \pi $ &0.450& 0.454  &0.214 &0.496\\
$ \iota $ &0.342&0.082 &0.052 &0.032\\
$  \rho^{\textsc{h}|\textsc{v}} $&0.476& 0.458& 0.290 & 0.464\\
$  \rho^{\textsc{ic}|\textsc{h}} $&0.682&0.940 & 0.072& 0.994\\
\hline
\end{tabular}
\end{table}

\section{Case-study: influenza during the 2017/18 season}\label{sec5}
The \ac{UKHSA}, formerly \ac{PHE}, routinely collects several sources of data to monitor influenza cases. Many of these datasets have been used separately to provide information on the transmission (\cite{birrell2011bayesian},  \cite{baguelin2013assessing}, \cite{corbella2018exploiting})  and severity \citep{presanis2014synthesising} of influenza; nevertheless, a joint analysis of all the sources has never been performed. Here a joint model is fitted to multiple data from the 2017/18 influenza season, with the aim of retrospectively characterising both severity and transmission, while appropriately attributing stochasticity to the various processes. Among other datasets, the \ac{USISS} is analysed and the methodology proposed in Section \ref{sec3} is used to jointly exploit hospitalisation and \ac{IC} unit admissions data. 
\subsection{Data}
The following datasets are included in the analysis:
\begin{itemize}
\item Confirmed influenza cases collected from week 40 of 2017 to week 20 of 2018 via the \ac{USISS} comprising the weekly count of \textit{\ac{IC} admissions} from, in principle, all trusts in England \citep{health2011sourcesa} and the weekly count of \textit{hospitalizations} at all levels of care in a stratified sentinel sample of the trusts \citep{health2011sourcesb};
\item Daily counts of\textit{ \acp{GP} consultations} for \ac{ILI} from a sample of \ac{GP} monitored  by EMIS \citep{harcourt2012use} or The Phoenix Partnership \citep{phoenix};
\item Observed proportion of respiratory swabs taken on a selected subset of \ac{GP} patients consulting for \ac{ILI} that test positive for influenza (\textit{virological} positivity, \cite{RCGPdata}). 
\item Results from cross-sectional \textit{serological} surveys that inform on the presence of antibodies in the population \citep{Andre}. 
\end{itemize}

\subsection{Model Specification}

\begin{figure}[h]
\centering
\includegraphics[scale=1, trim={5cm, 17cm, 5cm, 4cm}, clip]{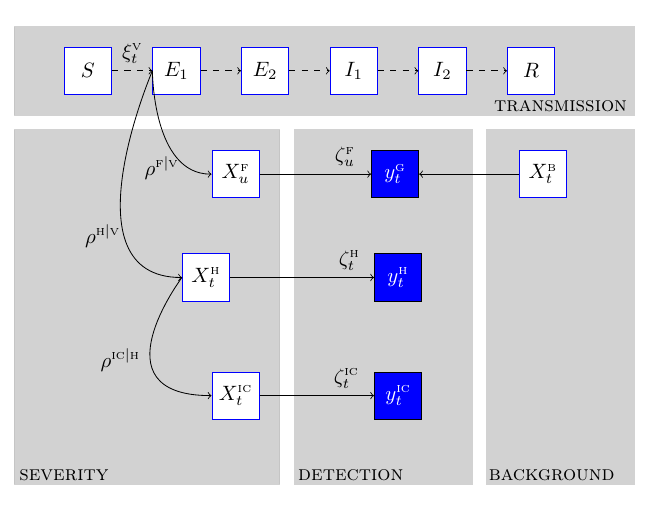}
\caption{
The transmission model classifies the population into susceptible ($ S $), exposed ($ E_1, E_2 $), infectious ($ I_1, I_2 $) and removed ($ R $). The severity model defines the occurrence of: mild flu cases $X^\textsc{f}_u $ with probability $\rho^{\textsc{f}|\textsc{v}} $, hospital cases $X^\textsc{h}_t $ with probability $\rho^{\textsc{h}|\textsc{t}} $, and of \ac{IC} cases $X^\textsc{ic}_t $ with probability $\rho^{\textsc{ic}|\textsc{h}} $. The detection process links cases to data defining the probability of reporting \ac{GP} consultations ($ \zeta_u^\textsc{g} $), hospitalizations ($ \zeta_t^\textsc{h} $) and \ac{IC} admissions ($ \zeta_t^\textsc{ic} $), respectively. The background process models the non-influenza \ac{ILI} cases, $ X^\textsc{b}_t $. }
\label{f5_2}
\end{figure}

Figure \ref{f5_2} provides an illustration of the data-generating processes which includes spread of the virus (transmission), probability of mild symptoms and severe outcomes (severity), background \ac{ILI} cases and detection. 

The model is specified in discrete time; intervals of length 1 day are denoted by the index $ u=1, 2, \dots $, while 1-week intervals are indexed by $ t=1, 2, \dots $ (e.g. daily \ac{GP} consultations are denoted by $ y^\textsc{g}_u $, while weekly hospitalizations are denoted by $ y^\textsc{h}_t $). The remaining notation is similar to that of Section \rev{\ref{sec3} and Section \ref{sec4}, with $ t $ and $ u $ indexing time in weeks and days, respectively.}
Some essential elements of the model are outlined here; the complete description of the model and the derivations of the data-distributions are provided in the Supplementary Information.

\subsubsection{Transmission and first severity layer}
Denote by $ \xi_{u}^\textsc{v}$ the number of new infections generated during day $ u $. A deterministic\rev{ $ SEEIIR $} transmission model, is assumed so that $ \xi_{u}^\textsc{v}$  is a function of the parameters $  \pi, \iota, \beta, \sigma, \gamma, \kappa  $ , representing the proportion of individuals initially immune; the proportion of initially infected/infectious individuals; the transmission rate; the rate of becoming infectious; the recovery rate; and the school-closure effect, respectively. \rev{Two compartments are used for each of the exposed and infectious state to allow for non-exponential visiting-times in those compartments \citep{keeling2011modeling}. }

The viral infection processes of individuals who will experience hospital admissions, $ X^{\textsc{h}|\textsc{v}_u} $, and influenza-related \ac{GP} consultations, $ X^{\textsc{f}|\textsc{v}_u} $, are assumed to follow a time non-homogeneous Poisson process, i.e.:
\begin{equation}
\begin{split}
\left(X^{\textsc{h}|\textsc{v}_u} \bigg|\xi^\textsc{v}_u,\rho^{\textsc{h}|\textsc{v}} \right)  &\sim \text{Pois}\left(\rho^{\textsc{h}|\textsc{v}}\cdot\xi^\textsc{v}_ut\right)  \qquad \text{for } u=0,1, \dots, U\\
\left(X^{\textsc{f}|\textsc{v}_u} \bigg|\xi^\textsc{v}_u,\rho^{\textsc{f}|\textsc{v}} \right) &\sim \text{Pois}\left(\rho^{\textsc{f}|\textsc{v}} \cdot\xi^\textsc{v}_u\right) \qquad \text{for } u=0,1, \dots,U
\end{split}
\label{eq5_16}
\end{equation}
with $ \rho^{\textsc{f}|\textsc{v}}$ and $ \rho^{\textsc{h}|\textsc{v}}$ denoting the probability of being visited by a \ac{GP} and being admitted to hospital, respectively. 

\subsubsection{\ac{GP}-consultations}
The inhomogeneous Poisson process of the people who will become mildly symptomatic and consult a \ac{GP}, $ X^{\textsc{f}|\textsc{v}_t} $ in Equation \eqref{eq5_16}, is the main component of the likelihood of the \ac{GP} consultations data. This process too is affected by delays that can be modelled by a discrete \ac{r.v.} modelling the days elapsing between infection and consulting a practice. 

In addition to these individuals, background, non-influenza cases appear in \ac{GP} consultation data; these endemic cases of other respiratory viruses and bacterial infections often follow a yearly seasonality, peaking around the same time as the seasonal influenza epidemic \citep{paul2008multivariate}. The background seasonality pattern is modelled by a weekly-varying  sine-cosine oscillation, similar to \cite{held2005statistical}. 

The general process describing the number of people consulting a \ac{GP} with \ac{ILI} symptoms is then obtained by adding the endemic and epidemic processes. Virological data are used to disentangle the proportion of the former process out of the total. 

Lastly, a Binomial emission is again chosen to model the observational process. However,  the probability of attending a \ac{GP} practice is subject to weekly fluctuations, caused by the weekend closure of \ac{GP} practices, hence a day-of the week distortion effect is included in the probability of detecting a \ac{GP} consultation.

\subsubsection{Hospitalization and \ac{IC} admissions}
A model for dependent data on hospitalizations and \ac{IC}-admissions data is illustrated in Section \ref{sec3}. Here too the joint likelihood is factorised in the marginal distribution of the \ac{IC} admissions $  Y_{1:T}^{\textsc{ic}} $, from the Poisson distribution of Equation \eqref{eq14}, and the distribution of the hospitalizations $ Y_{1:T}^{\textsc{h}} $ conditionally on \ac{IC} data approximated via \ac{MC} integration as proposed in Algorithm \ref{alg2}.

\subsection{Results}
The joint distribution of the unknown parameter vector (including transmission, severity, background and detection parameters) is derived via pseudo-marginal methods. Weakly-informative priors are assumed for most of the parameters of interest, all the prior distributions used and the estimated posteriors are reported in the Supplementary Information. 

\begin{figure}[!ht]
	\begin{subfigure}[c]{0.5\textwidth}
		\includegraphics[scale=0.36, page=1]{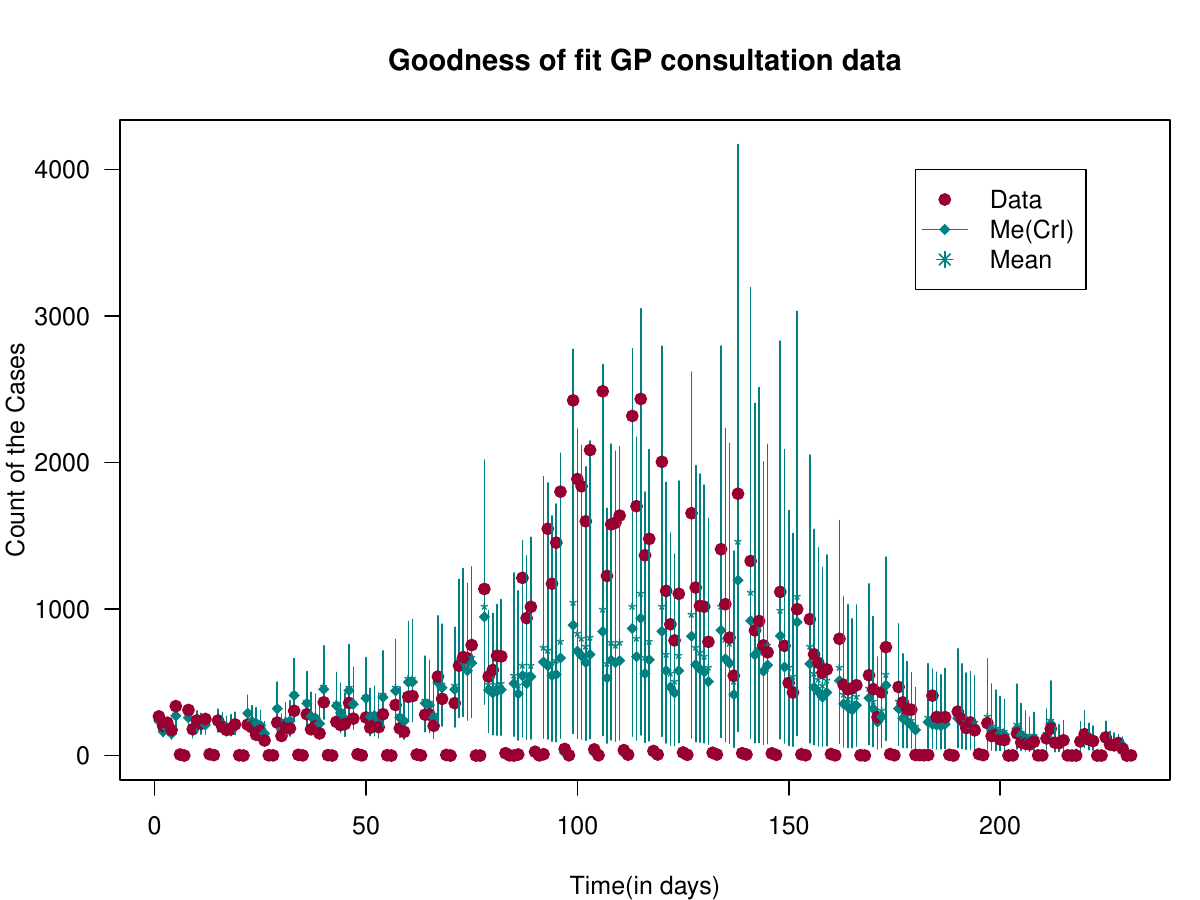}
		\caption{ }
	\end{subfigure}\hfill\begin{subfigure}[c]{0.5\textwidth}
	\includegraphics[scale=0.36, page=3]{GOF.pdf}
	\caption{ }
\end{subfigure}
	\begin{subfigure}[c]{0.5\textwidth}
	\includegraphics[scale=0.36, page=4]{GOF.pdf}
	\caption{ }
\end{subfigure}\hfill\begin{subfigure}[c]{0.5\textwidth}
	\includegraphics[scale=0.36, page=5]{GOF.pdf}
	\caption{ }
\end{subfigure}
\caption{Median and 95\% \acp{CrI} (green) for the posterior predicted distribution of: \ac{GP} data (a); virological data (b); hospitalizations (c) and \ac{IC} admissions (d). Red points are the observed data.}
\label{f6}	
\end{figure}

The model presents a fair fit to the data (Figure \ref{f6}) with all the observations being included in the 95\% \acp{CrI} of the posterior predictive distributions of the \ac{GP} consultation, virology and hospitalization data. The model, however, struggles to fit \ac{IC} data which might be in conflict with other sources of information in the model. These data, unlike \ac{GP} and hospital data, don't suggest a second peak in infection in the latest weeks of winter. 

\begin{figure}[h]
\begin{subfigure}[c]{7.56cm}
\centering
\includegraphics[scale=0.35]{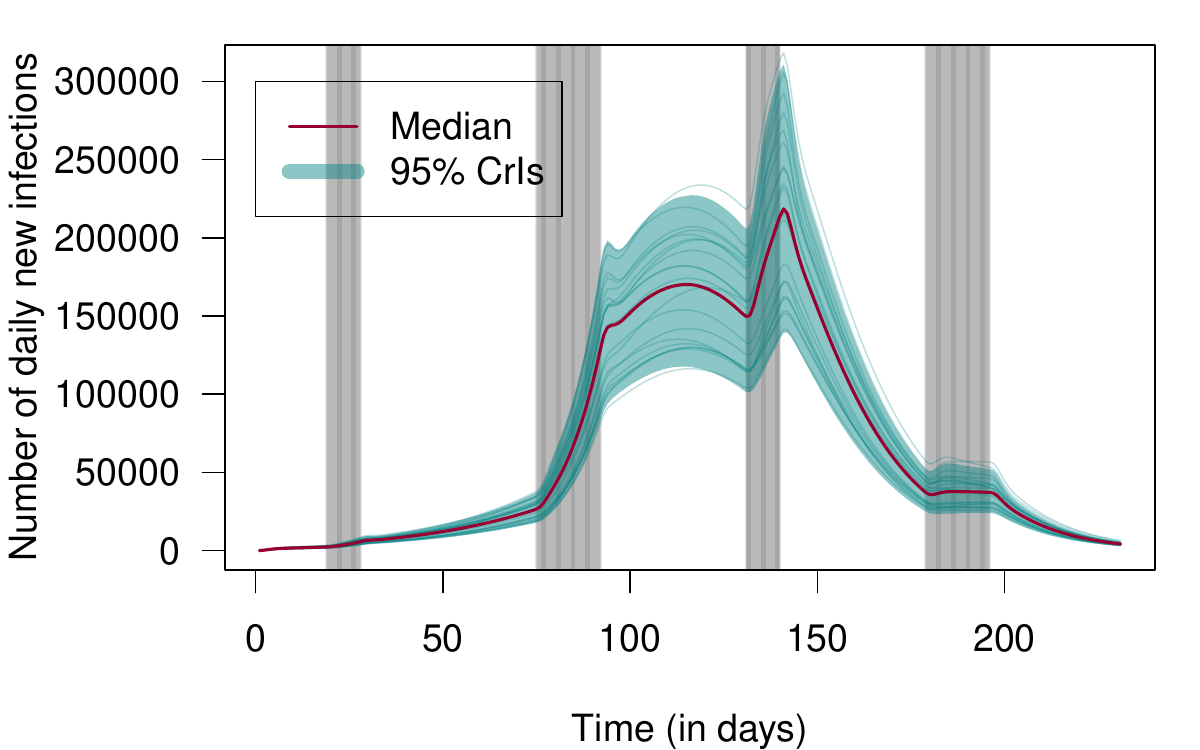}
\caption{Median (red) and 95\% \acp{CrI} (green) of the daily number of new infections. The grey areas corresponding to periods of lower-transmission (school holidays). 20 randomly selected trajectories are also plotted as thin green lines.}
\label{f7a}
\end{subfigure}\hfill\begin{subfigure}[c]{6.13cm}
\centering
\includegraphics[scale=0.33]{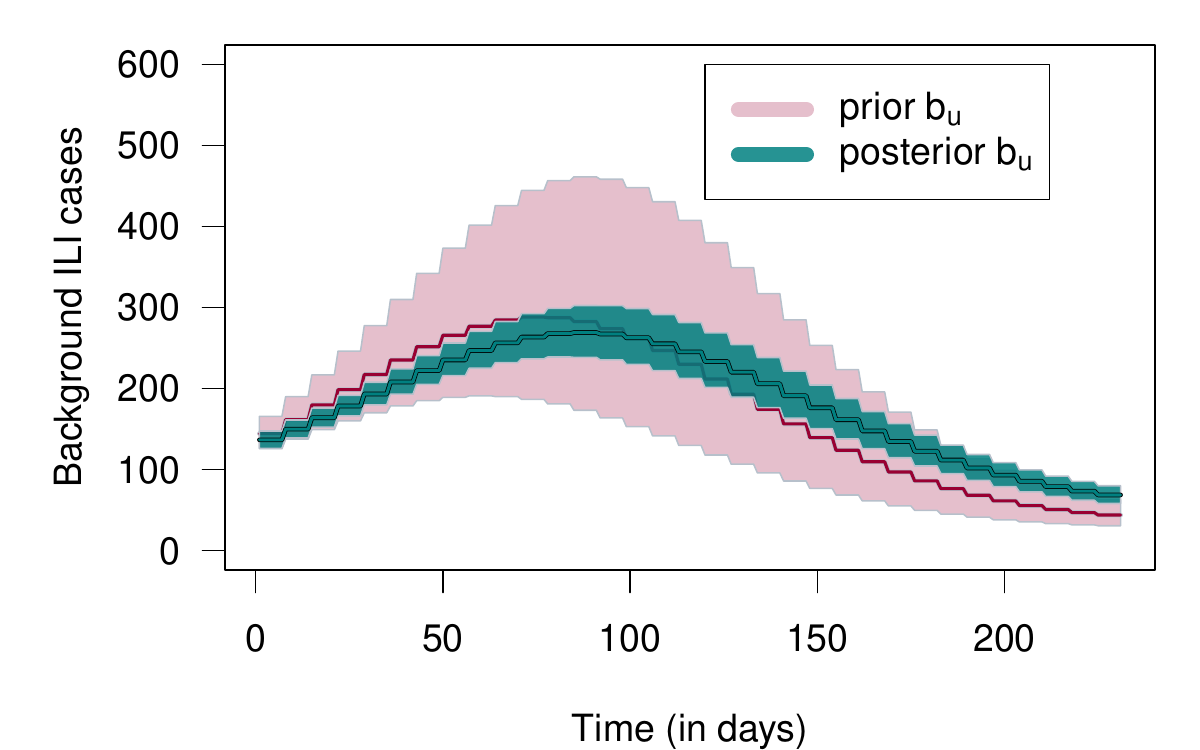}
\caption{Median (solid line) and 95\% \acp{CrI} (shaded area) of the prior (red) and posterior (green) for $ b_u $, the mean of the rate of the daily number of non-influenza \ac{ILI} \ac{GP} consultations.}
\label{f7b}
\end{subfigure}\\
\begin{subfigure}[c]{15cm}
\centering
\hspace*{-1.5cm}\includegraphics[scale=0.45, page=1]{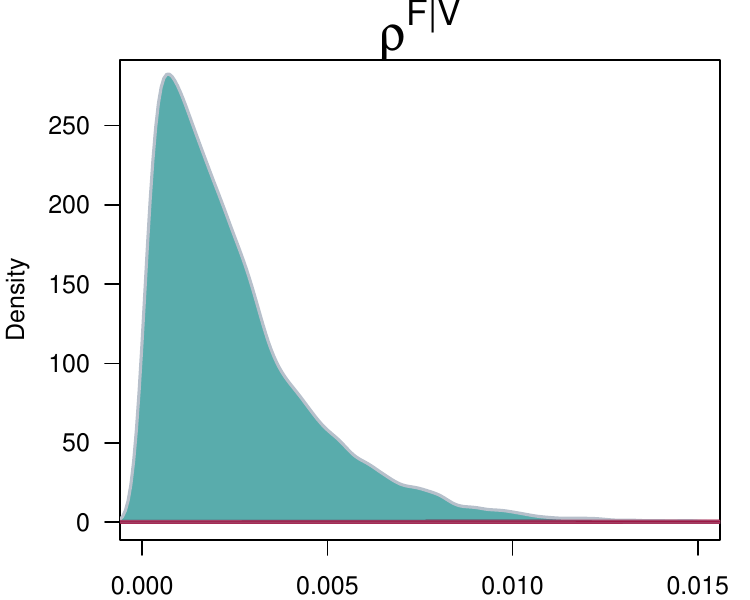}\includegraphics[scale=0.45, page=3]{fig6_3_4.pdf}\includegraphics[scale=0.45, page=4]{fig6_3_4.pdf}
\caption{Posterior (green), and prior (red) distribution of the severity parameters.}
\label{f7c}
\end{subfigure}
\caption{Posterior results from the joint analysis of 2017/18 data: transmission (a),  background (b), and some severity parameters (c).}
\label{f7}
\end{figure}

The model provides a comprehensive picture of influenza season 2017/18. Transmission, in terms of daily number of new infections is inferred and background \ac{ILI} cases are also described (Figure \ref{f7}). Moreover, all the key severity parameters of interest are well identified and informed by the joint use of the data, including the case-hospitalization risk (Median 0.0032, 95\% \ac{CrI} 0.0022-0.0049) and the hospital-\ac{IC} admission risk (Median 0.0667, 95\% \ac{CrI} 0.0574-0.078). 

The model in general presents many levels of sophistication. Firstly, \ac{ILI} cases attributable to influenza are disentangled from the background \ac{ILI}, which, while not being directly related to influenza, contributes to pressure on \ac{GP} practices. Secondly, the model accounts for peculiarities of the \ac{GP} data, such as the day-of-the-week variation, uncovering the true underlying transmission and severity process. Lastly and more importantly, exact inference from dependent data via \ac{MC} methods enables the simultaneous use of information from hospital and \ac{IC} surveillance. \rev{Crucially, these results were achieved only through the \textit{joint} use of all the data which together provide meaningful estimates of the severity and the transmission of influenza. When we attempted to fit similar transmission-severity models to each time series separately, results showed several issues including identifiability problems and poor model fit (see Section 2.5 of the Supplementary Information). }

\section{Discussion} \label{sec6}
 \rev{This work is set in the Bayesian framework, which provides a natural setup for the problem, allowing the fusion of information coming from multiple sources and from prior knowledge. }This paper contributes to the state-of-the-art literature on the analysis of epidemic data, from a methodological/computational perspective, providing new tools applicable to a wide class of problems that includes dependent data; and from an application perspective, providing innovative results on the analysis of influenza from multiple sources.
\subsection{Methodological and computational perspective}
Section \ref{sec2} contains a concise review of \acp{SSM} and their analysis, particularly focussed on the analysis of multiple epidemic data. There are other reviews of these methods in the literature (\cite{kantas2015particle, schon2018probabilistic}), some of which even target epidemic applications (\cite{mckinley2014simulation, dureau2013ssm, ionides2006inference}). However, they mainly address temporal dependences and, to the knowledge of the authors, little is known about the dependence across other domains. In the context of epidemic analysis, where temporal dependence can be easily overcome through realistic deterministic approximations of transmission dynamics, it is key to consider the sources of stochasticity and dependence linked to other processes (in this case the severity process) and its implications. 

Sections \ref{sec3} and \ref{sec4} embrace this challenge and highlight the possible complexities of such an analysis. Starting from a specific example, where there is dependence between data sources, an approach to the estimation of the static parameters of the system is proposed.
Key aspects such as overlaps between datasets and delays between consecutive events are then revealed to affect results. It is shown via simulation that simplifying assumptions, which assume independence among data and avoid expensive integration, lead to overconfident results and misleadingly narrow interval estimates. 

Our general and comprehensive way to set up epidemic models are paired with a specific estimation routine for their static parameters: pseudo-marginal methods. These methods are robust and simple to set up since they rely on the standard \ac{MH} algorithms \rev{ and, differently from other popular methods such as \ac{ABC}, they provide exact inference. Nevertheless,} they are expensive in two respects: firstly, a high number of simulations of the hidden states is required at each iteration of the static parameter to approximate its likelihood; and, secondly, the parameter space might be explored inefficiently due to their random walk behaviour. The first of these aspects might be addressed, for example, by converting to a \ac{DA} \ac{MCMC} \citep{o1999bayesian}, which only needs a single sample from the hidden states per \ac{MCMC} iteration. \ac{DA} approaches however, require tailored implementations for each different system considered, and therefore do not provide a general recipe applicable to a wide class of models. In terms of exploration of the parameter space, the unavailability of closed-form likelihood prevents the use of popular gradient based methods (e.g. \ac{HMC} \citep{neal2011mcmc} or the more recent \ac{PDMP} sampler \citep{bierkens2019zig}). Alternatively, a natural way to improve the estimation routine would be to consider \ac{SMC} methods on the parameter space, which in this context can be thought of as \ac{SMC}\textsuperscript{2} \citep{chopin2013smc2}.

\subsection{Application perspective}
The ability to estimate key transmission and severity parameters from epidemic data has never been more crucial. With the COVID-19 epidemic affecting almost every aspect of people's life, evidence-based decision-making has become an imperative aspect of public health policy. Uncertainty quantification around central estimates is key to understand both the extreme possible scenarios that could present in the near future and what further information is needed to increase knowledge and inform decisions. 

Section \ref{sec5} of the paper proposes a complex joint analysis of multiple data on influenza where both mild and severe cases inform transmission and severity parameters. The model presents multiple levels of sophistication including: the use of dependent data; the use of data from both non-confirmed and confirmed cases; the disentangling of background \ac{ILI} cases; the presence of delays between events. The key innovation of this model is that it matches a deterministic transmission dynamic with a stochastic severity and reporting process. In this sense it differentiates from other works that make joint use of multiple data but assume deterministic severity dynamics (e.g. \cite{keeling2020fitting}). 

\cite{shubin2016revealing} proposes a similar analysis of the 2009 pandemic in Finland. Even though his model presents many interesting aspects (e.g. the inclusion of both environmental and demographic stochasticity), unrealistic simplifications are made to allow for estimation in reasonable time, including the exclusion of an \textit{exposed} compartment, the discretization of time into extremely-large intervals (one week), and the assumption of no time elapsing between events of increasing severity. We consider our approach as a valid alternative to \cite{shubin2016revealing}, especially on a large population such as that of England, where the deterministic assumption is even more likely to hold. 

Going forward, the model could be extended in, at least, two interesting ways: firstly, population heterogeneity could be considered, through the formulation of an age-specific model that makes use of contact patterns between age groups; and secondly, more overdispersion could be allowed in the severity process by, for example, replacing Poisson \acp{r.v.} with Negative Binomial \acp{r.v.}. These changes would almost certainly improve the fitting of the model to data.

\rev{Data collected during the COVID-19 epidemic could provide a further context where to test our semi-stochastic model and our inferential routine that accounts for dependent data. Similarly to influenza, the COVID-19 pandemic presented with high numbers of infected individuals, allowing the approximation of the transmission dynamics to a deterministic system. Nevertheless, the number of people presenting with severe symptoms and the timing of accessing healthcare facilities was highly stochastic. While, in the UK as in many other counties, all the data were collated by a single institution, data might have overlaps and conflicts, hence our methodology would provide a proper allocation of the noise and dependencies intrinsic to the system.  }

\begin{acronym}
\acro{SSM}{state-space model}
\acro{r.v.}{random variable}
\acro{POMP}{partially observed Markov process}
\acrodefplural{POMP}{partially observed Markov processes}
\acro{HMM}{hidden Markov model}
\acro{BPF}{bootstrap particle filter}
\acro{MC}{Monte Carlo}
\acro{HMC}{Hamiltonian Monte Carlo}
\acro{SMC}{sequential Monte Carlo}
\acro{ABC}{Approximate Bayesian Computation}
\acro{MH}{Metropolis Hastings}
\acro{MCMC}{Monte Carlo Markov chain}
\acro{GIMH}{grouped independence Metropolis Hastings}
\acro{MCWM}{Monte Carlo within Metropolis}
\acro{DAG}{directed acyclic graph}
\acro{USISS}{UK Severe Influenza Surveillance System}
\acro{NHS}{National Health Service}
\acro{IC}{Intensive Care}
\acro{CrI}{Credible Interval}
\acro{PHE}{Public Health England}
\acro{ILI}{influenza-like illness}
\acro{GP}{General Practitioner}
\acro{RCGP}{Royal College of General Practitioners}
\acro{ONS}{Office of National Statistics}
\acro{UKHSA}{UK Health Security Agency}
\acro{DA}{data-augmented}
\acro{PDMP}{Piecewise Deterministic Markov Process}
\end{acronym}

\section*{Acknowledgements}
The author would like to thank colleagues at Public Health England, particularly Dr Richard Pebody, Dr Andre Charlett and Dr Nikolaos Panagiotopoulos, for providing the data and prior information for the analysis in Section 6.
We are grateful to Dr Mi\rev{k}hail Shubin for insightful discussions on the earlier work on dependent data streams. Lastly, Dr Trevelyan J McKinley and Dr Paul Kirk provided useful feedback on an earlier version of this work for which we are thankful.

AC is supported by Bayes4Health EPSRC Grant EP/R018561/1 \rev{and by the Royal Society (Dimension Supercharged Projective Sampling project)}. All four authors were supported  by the MRC, programme grant MC\_UU\_00002/11.

\begin{supplement}
\stitle{Supplementary Information}
\sdescription{This report contains further work on the study of dependence between data and supplementary information to the study of Section \ref{sec5}. The former includes further comments on SSM inference, the derivations of Algorithm \ref{alg2} and the results form the simulation study of Section \ref{sec4}. The latter includes extensive explanation of the model, inferential methods and results of the analysis.}
\end{supplement}
\begin{supplement}
\stitle{Supplementary code and data}
\sdescription{Code for the simulation study in Section \ref{sec4} and the analysis of Section \ref{sec5} are available at \href{https://github.com/alicecorbella/EpiDependentData}{https://github.com/alicecorbella/EpiDependentData}. The data analysed in Section \ref{sec5} are based on routine health-care data, which cannot be made available to others by the study authors. Requests to access these non-publicly available data are handled by the \href{https://www.gov.uk/government/publications/accessing-public-health-england-data/about-the-phe-odr-and-accessing-data}{Public Health England Office for Data Release} and its successor at the \href{https://www.gov.uk/government/publications/accessing-ukhsa-protected-data}{UK Health Security Agency}.}
\end{supplement}

\bibliographystyle{imsart-nameyear} 
\bibliography{bibliography.bib}

\end{document}